\documentclass[pra,twocolumn,superscriptaddress,aps,showpacs,amsmath,amssymb,nofootinbib]{revtex4-1}

\usepackage{epsfig,graphicx}
\usepackage{color}
\usepackage{amssymb}
\usepackage{amstext}
\usepackage{amsthm}
\usepackage{amsfonts}
\usepackage{latexsym}
\usepackage{cancel}
\usepackage{epstopdf}
\usepackage{array}
\usepackage{xfrac}

\def\beq{\begin{equation}}{\it}
\def\eeq{\end{equation}}
\def\beqa{\begin{eqnarray}}{\it}
\def\eeqa{\end{eqnarray}}

\begin{document}

\title{Dynamic structure function of two interacting atoms in 1D}
\author{David Ledesma}
\address{Departament de F\'isica Qu\`antica i Astrof\'isica,\\
Facultat de F\'{\i}sica, Universitat de Barcelona, E--08028 Barcelona, Spain}

\author{Alejandro Romero-Ros} 
\address{Departament de F\'isica Qu\`antica i Astrof\'isica,\\
Facultat de F\'{\i}sica, Universitat de Barcelona, E--08028 Barcelona, Spain}
\address{Zentrum f\"ur Optische Quantentechnologien,\\
Universit\"at Hamburg, Luruper Chaussee 149, 22761 Hamburg, Germany}

\author{Artur Polls}
\address{Departament de F\'isica Qu\`antica i Astrof\'isica,\\
Facultat de F\'{\i}sica, Universitat de Barcelona, E--08028 Barcelona, Spain}
\address{Institut de Ci\`encies del Cosmos, 
Universitat de Barcelona, IEEC-UB, Mart\'i i Franqu\`es 1, E--08028
Barcelona, Spain}

\author{Bruno Juli\'a-D\'iaz}
\address{Departament de F\'isica Qu\`antica i Astrof\'isica,\\
Facultat de F\'{\i}sica, Universitat de Barcelona, E--08028 Barcelona, Spain}
\address{Institut de Ci\`encies del Cosmos, 
Universitat de Barcelona, IEEC-UB, Mart\'i i Franqu\`es 1, E--08028
Barcelona, Spain}
\address{ICFO-Institut de Ci\`encies Fot\`oniques, Parc Mediterrani 
de la Tecnologia, E--08860 Barcelona, Spain}

\begin{abstract}
We consider two atoms trapped in a one-dimensional harmonic 
oscillator potential interacting through a contact interaction. We study 
the transition from the non-interacting to the strongly interacting 
Tonks-Girardeau state, as the interaction is varied from zero to infinitely 
large repulsive values. The dynamic structure function is computed by 
means of direct diagonalization calculations with a finite number of 
single particle modes. The response of the system against a monopolar 
perturbation is characterized by the moments of the dynamic structure 
function which are explicitly calculated from the dynamic structure 
function and by means of sum rules. 
\end{abstract}

\maketitle

\section{Introduction} 
\label{intro}

Recent experiments with few ultracold atoms trapped in harmonic potentials have 
opened the way to understand how many-body quantum mechanical correlations build 
in few-body systems~\cite{2010XeOE, 2011SerwaneScience,2012ZurnPRL,2013WenzScience,
2013BourgainPRA,2014PaganoNatphys,2014Murmann,regal15,esslinger2003,nagerl2009,bouchoule2014}.
Such systems, if further confined to a 1D geometry, offer an appropriate playground 
to study the interplay between quantum many-body correlations and statistics. In 
the present paper, we concentrate in the case of bosons confined in a 1D harmonic 
oscillator potential and  study the behavior of the ground state and collective 
excitations ~\cite{esslinger2003,nagerl2009,bouchoule2014}.

The simplest example is that of two particles trapped in a harmonic oscillator 
potential~\cite{Busch98}. For the case of a contact interaction potential, which 
is realistic for most  ultracold atomic systems~\cite{BECtrapped}, two noteworthy 
limits are well known. In absence of interactions the two atoms populate the lowest 
energy single particle state, producing the minimal version of a Bose-Einstein 
condensed state. In the infinite interaction limit the two bosons resemble in 
many ways two spinless non-interacting fermions, producing the so called 
Tonks-Girardeau (TG) limit~\cite{girardeau}. The behavior of the two body system in 
the two limits is completely different \cite{idziaszek2006}, reflecting the nature of the quantum 
correlations between the atoms~\cite{excitations,zollner2006,brouzos2012,kehrberger2018}. This 
behavior can be probed by exciting the trapped two-atom system with the monopolar 
excitation operator and observing the response.
The first collective excitation of the trapped system has the same excitation energy 
of $2\hbar\omega$ for both the non-interacting and infinitely interacting limits, 
being $\omega$ the trap frequency~\cite{breathing}. 
This mode has actually two contributions. The first arises from the excitation of the center-of-mass 
mode. This contribution remains unchanged as we increase the atom-atom interaction 
and solely depends on the trapping potential. The second contribution is more 
intricate and corresponds to the first excited mode of the relative motion, e.g. 
a breathing mode.
The excitation energy of this mode changes as we vary the interaction strength.
Starting at $2\hbar\omega$ in the non-interacting case, it decreases as the atom-atom 
interaction strength is increased until it reaches a minimum. This increase of the 
interaction also builds up strong correlations between the two atoms.
Past the minimum, the excitation energy increases with the interaction towards its 
limiting value of $2\hbar\omega$, i.e. the same value as the non-interacting case.
This reentrance behavior of the breathing mode has been experimentally 
observed in the case of cesium atoms trapped in very elongated independent tubes 
(one-dimensional systems) with each tube containing between 8 to 25 atoms at 
their center~\cite{nagerl2009}.

The breathing mode can be excited by varying the trapping potential. That is, for 
instance, it can be excited by first preparing the system in the ground state 
corresponding to $\omega$ and then suddenly quenching the trapping frequency. In 
Refs.~\cite{schmelcher,phyz2018} the excitations of the system were studied following this 
procedure. The excitation energies were obtained from the Fourier analysis of the 
beating in the system after the quench. Direct diagonalization methods were used 
in Ref.~\cite{haque}, where the continuum was  studied as a limiting case for 
discrete Bose-Hubbard models. On the other hand, in Ref.~\cite{breathing} quantum 
Monte-Carlo calculations were performed for systems of up to $N=25$ atoms. The 
method utilized to investigate the breathing mode was to determine the average 
excitation energy by means of sum rules~\cite{joan,menotti2002,abraham2014}, i.e, by computing 
the expectation value of certain operators in the ground state of the system. 

In this article we compute the dynamic structure function (DSF) and study 
the exact response of the system associated to a monopolar excitation as we vary 
the atom-atom interaction, which is taken always repulsive.
We compare the results obtained by means of the sum rules~\cite{joan,excitations,menotti2002} 
previously considered~\cite{breathing} with a direct computation integrating the 
structure function. The article is organized in the following way. In 
Sec.~\ref{sec1} we describe the two particle Hamiltonian and explain our 
diagonalization procedure. In Sec.~\ref{dynstrufunc} we present the dynamic 
structure function computed at several interaction strengths. Explicit expressions 
for the sum rules applied to our system are given in Sec.~\ref{sumrules}. 
In Sec.~\ref{conclusions} we provide a summary and the main conclusions 
of our work. Finally, there is an appendix with some technical details 
concerning the contribution of the center-of-mass.

\section{Hamiltonian}
\label{sec1}

The Hamiltonian for two bosons trapped in a one dimensional harmonic 
oscillator (h.o.) and interacting through a contact potential reads, 
\beqa
H &=& - \frac {\hbar^2}{2m} \frac {d^2}{dx_1^2} 
- \frac {\hbar^2}{2m} \frac {d^2}{dx_2^2}
+ \frac {1}{2} m \omega^2 x_1^2 + \frac {1}{2} m \omega^2 x_2^2 \nonumber\\
&& 
+g\,\delta(x_1-x_2) \,.
\eeqa
As customary, we can define the center-of-mass (c.m.) and relative coordinates 
as, $X= (x_1 +  x_2)/2$ and $x_{\rm r}= x_1-x_2$. This change of variables 
allows us to decompose the Hamiltonian in two pieces ,
$H= H_{\rm c.m.}+ H_{\rm r}$, with, 
\beqa
H_{\rm c.m.} &=& -\frac {\hbar^2}{2 M} \frac {d^2}{dX^2} 
            + \frac {1}{2} M \omega^2 X^2 \, ,
\nonumber\\
H_{\rm r}      &=& -\frac {\hbar^2}{2 \mu } \frac {d^2}{dx_r^2} 
           + \frac {1}{2} \mu \omega^2 x_r^2 
+g \ \delta(x_r) \,.
\eeqa
where $M=2 m$ is the total mass and $\mu= m/2$ is the reduced mass. 
The frequencies associated to both Hamiltonians are the same.
However, the oscillator lengths are different. Now, it is convenient to express 
the Hamiltonians in the oscillator units of the original Hamiltonian: 
\beqa
H_{\rm c.m.} &=& -\frac{1}{4} \frac {d^2}{d X^2} + X^2 \,,
\nonumber \\
H_{\rm r}     &=& - \frac {d^2}{dx_r^2} + \frac {1}{4} x_r^2 + g \ \delta(x_r) \,, \label{eq:3}
\eeqa
where all quantities have been properly re-scaled  using the harmonic oscillator 
units. From now on, we will refer to the $H_{\rm r}$ without the interaction term, 
$g\delta(x_r)$, as $H_{\rm r}^{(0)}$.

As the Hamiltonian splits in two commuting ones, the wave functions associated 
to the c.m. motion and the relative motion factorize, 
\begin{equation}
\Psi(X,x_r) = \phi(X) \, \psi(x_r) \,.
\end{equation}
$H_{\rm c.m.}$ is a harmonic oscillator Hamiltonian with the spectrum 
$E_k^{c.m.} = k+1/2$ where $k$ is the number of quantums associated 
to the c.m. motion.

\begin{figure}[t]
\centering
\includegraphics[width=1. \columnwidth]{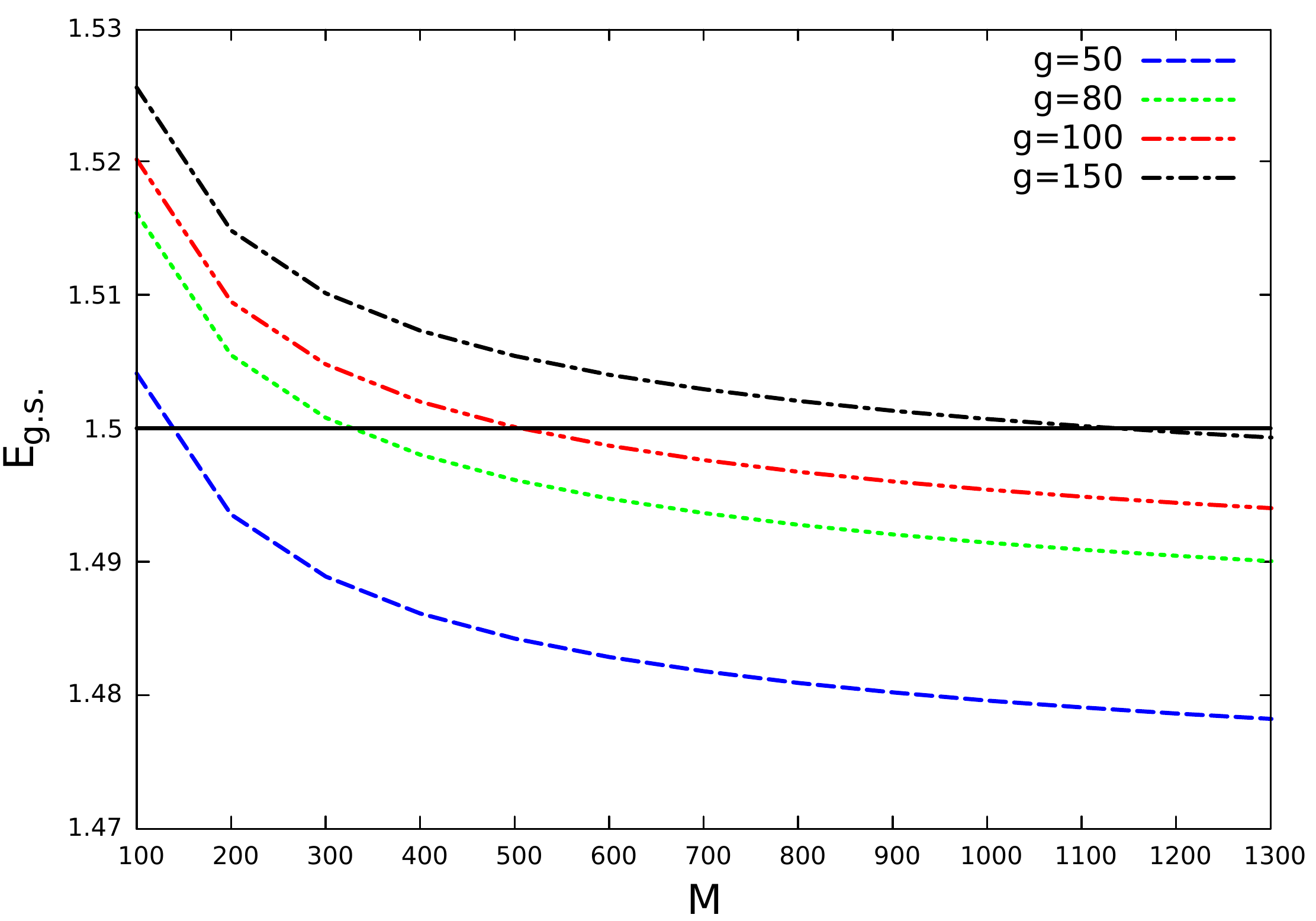}
\caption{Ground state energy (in h.o. units) of $H_{\rm r}$ as a function 
of the number of modes (the dimension of the space used to diagonalize 
the Hamiltonian, $M$) for several values of the interaction strength $g$.}  
\label{fig:conv}
\end{figure}

To diagonalize the relative part we use the following procedure. We choose a finite 
number of non-interacting energy eigenstates, $\{\psi_0(x_r),\psi_2(x_r),\dots,\psi_{2M}(x_r)\}$, 
corresponding to $H_{\rm r}^{(0)}$.
Notice that, in order to fulfil the symmetry requirements of the two-boson wave function, 
we only consider the modes associated to even functions. Hence, the matrix 
elements of $H_{\rm r}$ are: 
\beqa
\langle \Psi_{2m} | H_r |\Psi_{2n}\rangle &=& \left ( \frac {1}{2} + 2n \right )
 \delta_{m,n}  \\
& + & 
g \ \int \ dx_r \  \Psi_{2n} (x_r) \Psi_{2m} (x_r) \delta(x_r) \nonumber \\ 
&=& \left ( \frac {1}{2} + 2n \right ) \delta_{m,n} + g \Psi_{2n}(0)\Psi_{2m}(0) \nonumber \,.
\eeqa
Diagonalizing the truncated Hamiltonian we get approximate solutions, whose 
eigenvalues are upper bounds of the corresponding exact solutions, 
\beq
H_r \tilde \Psi_\ell = \tilde E^{\rm (r)}_\ell \tilde \Psi_\ell \,.
\eeq
In practice we use a huge number of modes, $M$ up to 1500, to guarantee the convergence 
of the calculations. In fact, in Fig.~\ref{fig:conv} we  report the dependence of the 
ground state energy on  the number of modes used to diagonalize $H_{\rm{r}}$ for different 
values of the interaction strength $g$. Obviously, the energy decreases as the number of 
modes increases reaching in all cases a limit saturating value with the number of modes 
which defines an  asymptotic value. The horizontal line at $E_{g.s.}=3/2$ corresponds to 
the ground state energy of $H_{\rm{r}}$ in the limit $g\to \infty$. For any finite 
value of $g$ one can always find a number of modes ($M_{crit}$) such that for a larger 
number of modes the energy will be always below $3/2$. The $M_{crit}$ becomes larger as 
$g$ increases. Therefore, one can conclude that when $g\to \infty$ the ground state 
energy of $H_{\rm{r}}$, as a function of the number of modes, approaches $3/2$ from below.    

\begin{figure}[t]
\centering
\includegraphics[width=1\columnwidth, angle=0]{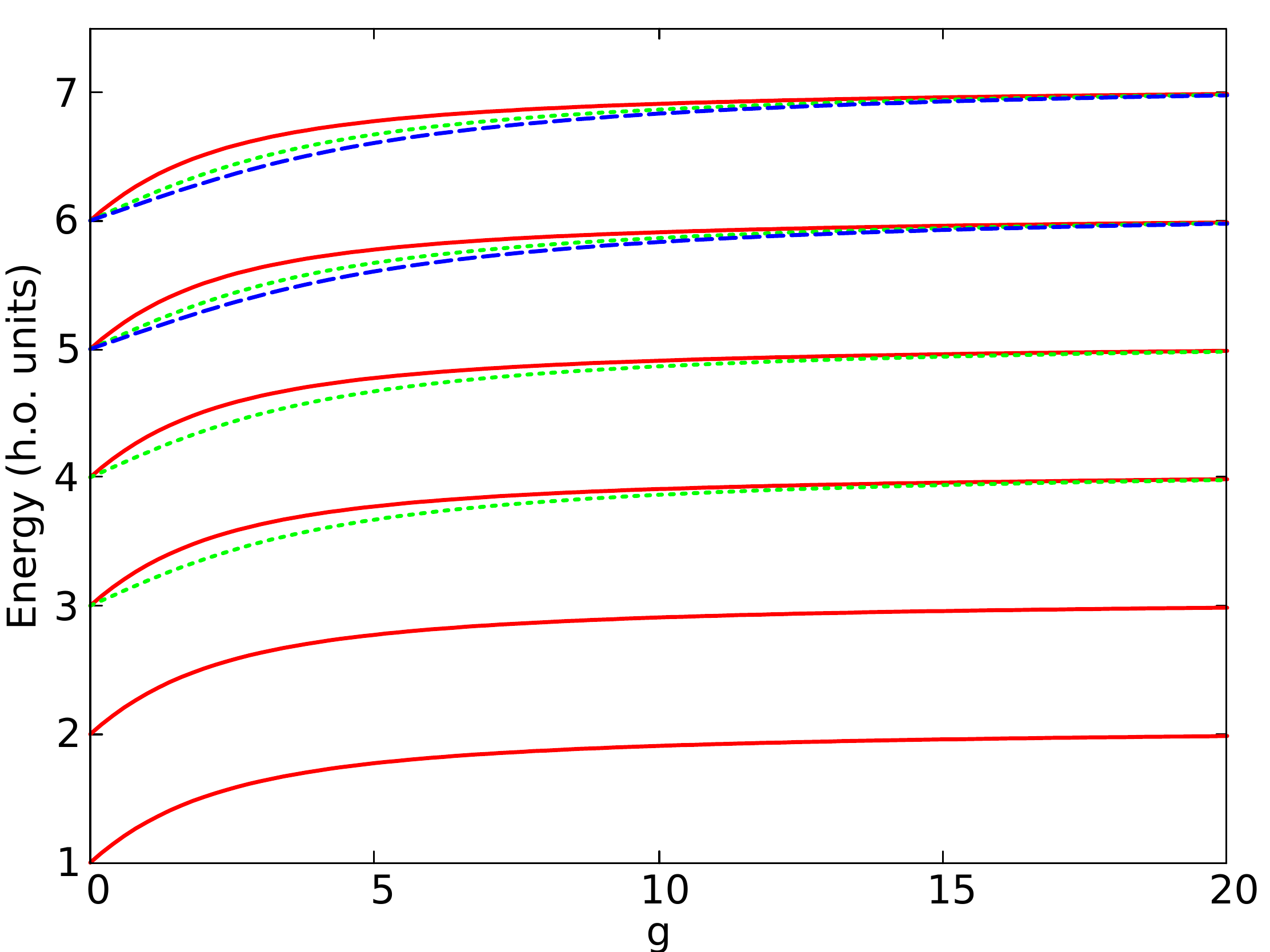}
\caption{Lowest energy levels of the spectrum for the two bosons system, for 
a range of interaction strength $g=[0,20]$. The red lines correspond to the 
ground state  of $H_{\rm r}$, whereas the green and blue lines correspond to 
1st and 2nd relative excitations, respectively. These relative Hamiltonian 
states are also combined with c.m. excitations, which result in parallel 
curves which are shown with the same color for the same relative Hamiltonian 
state. For instance, the lowest state takes into account the energy of the 
ground state of the c.m., equal to $1/2$ and independent of the strength 
of the interaction, and the ground state energy of $H_{\rm r}$ which is $1/2$ 
for $g=0$ and approaches $3/2$ when $g \rightarrow \infty$.}
\label{fig:spec}
\end{figure}

For the case of two bosons trapped in a harmonic oscillator potential, we 
could have used the exact solutions of Ref.~\cite{Busch98}. Instead, we have 
decided to present a procedure which, although more tedious for this precise 
case, is more general and can readily be applied to other single particle 
trapping potential besides the harmonic oscillator or also to other interactions 
by a proper computation of the matrix elements. 
Our results coincide, when applicable, with the exact solutions of Ref.~\cite{Busch98}.

Diagonalization methods have also been used for systems with larger number 
of particles. However,the separation of the problem in c.m. and 
intrinsic coordinates is more difficult to implement. In those cases, the 
single-particle wave functions  associated to the harmonic trap in the laboratory 
system are used to build the many-body states of the Fock space. The proper 
situation of the c.m. excitation modes serves as a test of the size 
of the Fock space used to diagonalize the Hamiltonian~\cite{garcia.march2013,bongs2007,schmelcher}.

The full spectrum of the problem will be obtained as, 
\beq
E_{k,\ell}= E^{(\rm c.m.)}_k + \tilde E^{(\rm r)}_\ell \,.
\eeq
In Fig.~\ref{fig:spec} we depict the lowest part of the spectrum as a function 
of $g$. One can see how the eigenvalues evolve from the non-interacting bosonic 
system to those of a free fermionic system for $g\to \infty$. 
At $g=0$, the ground state of $H$ is non-degenerate. It is built with both atoms in the 
ground state of the harmonic oscillator single-particle Hamiltonian or, which 
is the same, as the product of the ground states of $H_{\rm {c.m.}}$ and $H_{\rm r}$. 
Thus, both descriptions provide the same energy, $E=1$. However, in the
$g\rightarrow \infty$ limit, the ground state of the two-body system is given by the 
absolute value of the Slater determinant built with one atom in the ground state 
and the other in the first excited state of the harmonic oscillator single particle 
Hamiltonian. Therefore, in this limit, the ground state energy (2 in h.o. units) is 
the sum of $1/2$ and $3/2$ single-particle energies, and the contribution of the 
contact interaction is zero. When one does the decomposition in $H_{\rm {c.m.}}$ and 
$H_{\rm{r}}$ the ground state in the $g \rightarrow \infty $ is achieved by taking 
the c.m. in the ground state ($\phi_0(X)$), i.e. in the lowest c.m. level
with energy $1/2$, and taking the relative wave function to be the absolute value 
of the first excited state, $\psi_1(x_r)$ of $H_{\rm{r}}^{(0)}$ with energy $3/2$. Notice 
that $\psi_1(x_r)$ is an odd function and therefore it would be necessary to use an 
infinite basis of even functions, which are the ones that we have used in the 
diagonalization of $H_r$ to respect the symmetry requirements for bosons, to 
asymptotically approach the exact eigenstate. The lowest excited state of the system 
corresponds to the  first excitation of $H_{\rm {c.m.}}$, with energy $3/2$ and to 
the relative ground state along $g$. Thus, the first excited state has a constant 
excitation energy independent of $g$. Actually, the excitations of the c.m. show 
all the way up resulting in a set of curves parallel to the ground state curve, all 
depicted with red colour in Fig.~\ref{fig:spec}. Notice that the first excited state 
at $g=0$ can also be described as a properly symmetrized wave function with an atom 
in the ground state (energy 1/2)  and the other in the first excited sate (energy 3/2) 
of the single-particle harmonic oscillator Hamiltonian.

The next excitation consists of two states: one corresponds to $2$ quantums 
of excitation of the c.m. motion combined with the ground state of $H_{\rm r}$ (red-line);
the other one describes the state where the c.m. is in the ground state 
and the relative motion is in the first excited state of positive parity of $H_r$ (green line). 
The two lines coincide at $g=0$, where  the energy level has degeneracy 2.
Then, as $g$ increases, the degeneracy is broken (green line) and the energy of 
this state is always below the energy of the state with two quantums of excitation 
of the c.m. At $g \rightarrow \infty$ the two states become again degenerated with total energy 4. 

\begin{figure}[t]
\centering
\includegraphics[width=1.0\columnwidth, angle=-0]{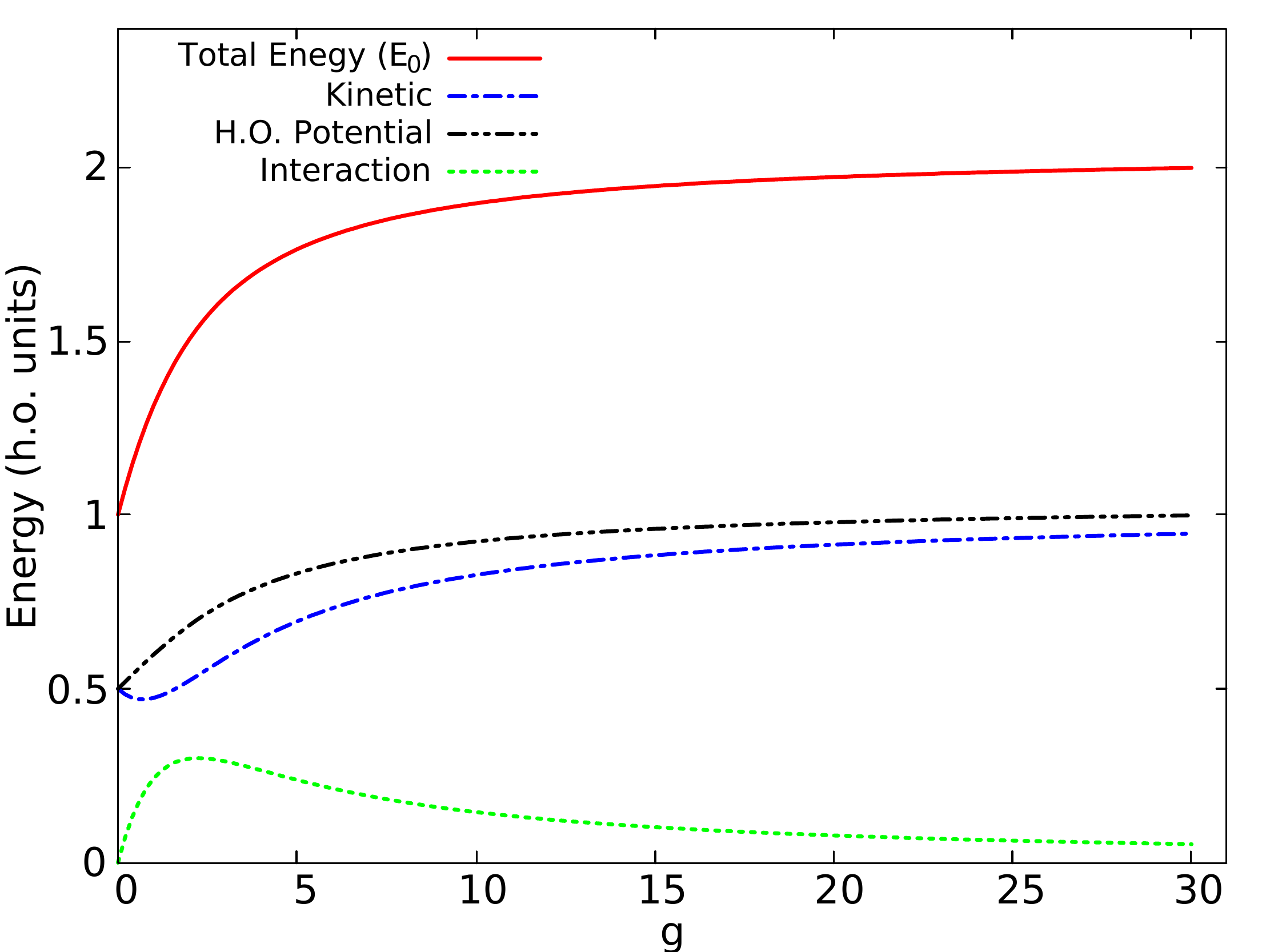}
\caption{Different contributors  to the ground state energy: $V_{h.o.}$, $E_{\rm kin}$ and $V_{int}$, 
as functions of the interaction strength $g$. The total ground state energy is shown 
in red, $E_{\rm kin}$ in blue, $V_{h.o.}$ in black and $V_{int}$ in green.} 
\label{fig:fign}
\end{figure}

The transition from the BEC regime to the TG regime as the interaction strength 
is increased is shown in Fig.~\ref{fig:fign}. There, we report as 
a function of $g$ the total ground state energy (red line) decomposed in  
kinetic energy $E_{\rm kin}$ (blue line), harmonic oscillator potential energy 
$V_{\rm h.o.}$ (black line) and interaction energy $V_{int}$ (green line). 
We observe that $V_{\rm h.o.}$ increases monotonically from $V_{\rm h.o.}(g=0)=1/2$ up to 
$V_{\rm h.o.}(g\to\infty) = 1$, while the kinetic energy starts at $E_{\rm kin}(g=0)= 1/2$ 
and goes through a minimum before before growing up to $E_{\rm kin}(g\to\infty)= 1$.
Notice that both $E_{\rm kin}$ and $V_{\rm h.o.}$ contain a constant contribution 
of the c.m. equal to 1/4. We also observe how the interaction energy increases 
with $g$ until it reaches a maximum around $g\simeq 2$. For $g\gtrsim 2$ the 
behavior changes completely and, despite the increasing atom-atom interaction 
strength, the interaction energy of the ground state decreases monotonically 
as $g\to \infty$. Notice that, when $g \rightarrow \infty$, the exact wave function 
for the relative motion, built as $| \psi_1(x_r) |$, does not feel the interaction.
In fact, $V_{int}(g=0)=0$ and $V_{int}(g\to\infty)=0$. This is a clear consequence 
of the formation of correlations in the system which avoid the contact of the 
two bosons.

Moreover, the virial theorem,
\begin{equation}
- 2 V_{\rm h.o.} + 2 E_{\rm kin} + V_{int} =0 \, ,
\end{equation}
is exactly fulfilled for any value of $g$, indicating a good convergence of the energy results with the number 
of modes. The virial relation, together with the fact that $E=E_{\rm kin}+V_{\rm h.o.}+V_{int}$, allows one 
to write
\begin{equation}
	E_{\rm kin} = \frac {1}{2} E - \frac {3}{4} V_{int} \,,  \qquad
	V_{\rm h.o.} = \frac {1}{2} E - \frac {1}{4}V_{int} \,.
\end{equation}	
Therefore, we conclude that $V_{\rm h.o.} \ge E_{\rm kin}$, for any value of $g$.
In particular, since $V_{int}=0$ at $g=0$, $E_{\rm kin}= V_{\rm h.o.}= E/2= 1/2$. 
The same happens at $g \rightarrow \infty$ when $V_{int}=0$, but now with $E/2=1$.

In order to understand the behaviour of the different pieces of the energy for 
very small $g$, we can combine a first order perturbation theory with the virial theorem. The correction to the 
energy is provided by the expectation value 
$V_{int,0}= \langle \phi_r^{(0)}| g \delta(x_r) | \phi_r^{(0)} \rangle = g \phi_r^{(0)}(0) \phi_r^{(0)}(0) = g (1/(2 \pi))^{1/2}$,  where $| \phi_r^{(0)}\rangle$ is the ground state 
of $H_{\rm r}^{(0)}$. Therefore, $V_{int,0}$ increases linearly with $g$. Using the 
previous relations for $E_{\rm kin}$ and $V_{\rm h.o.}$,
\begin{equation}
E_{\rm kin}= \frac {1}{2} - \frac {1}{4} V_{int,0} 
= \frac {1}{2} - \frac {1}{4} g \left ( \frac {1}{2\pi} \right )^{1/2} \, ,
\end{equation}
\begin{equation}
V_{\rm h.o.} = \frac {1}{2} + \frac {1}{4} V_{int,0} = \frac {1}{2} 
+ \frac {1}{4} g \left ( \frac {1}{2\pi} \right )^{1/2} \,.
\end{equation}
These expressions agree well with the low $g$ regime depicted in 
Fig.~\ref{fig:fign}.

\section{Dynamic structure function} 
\label{dynstrufunc}

\begin{figure}[t]
\centering
\includegraphics[width=1\columnwidth, angle=-0]{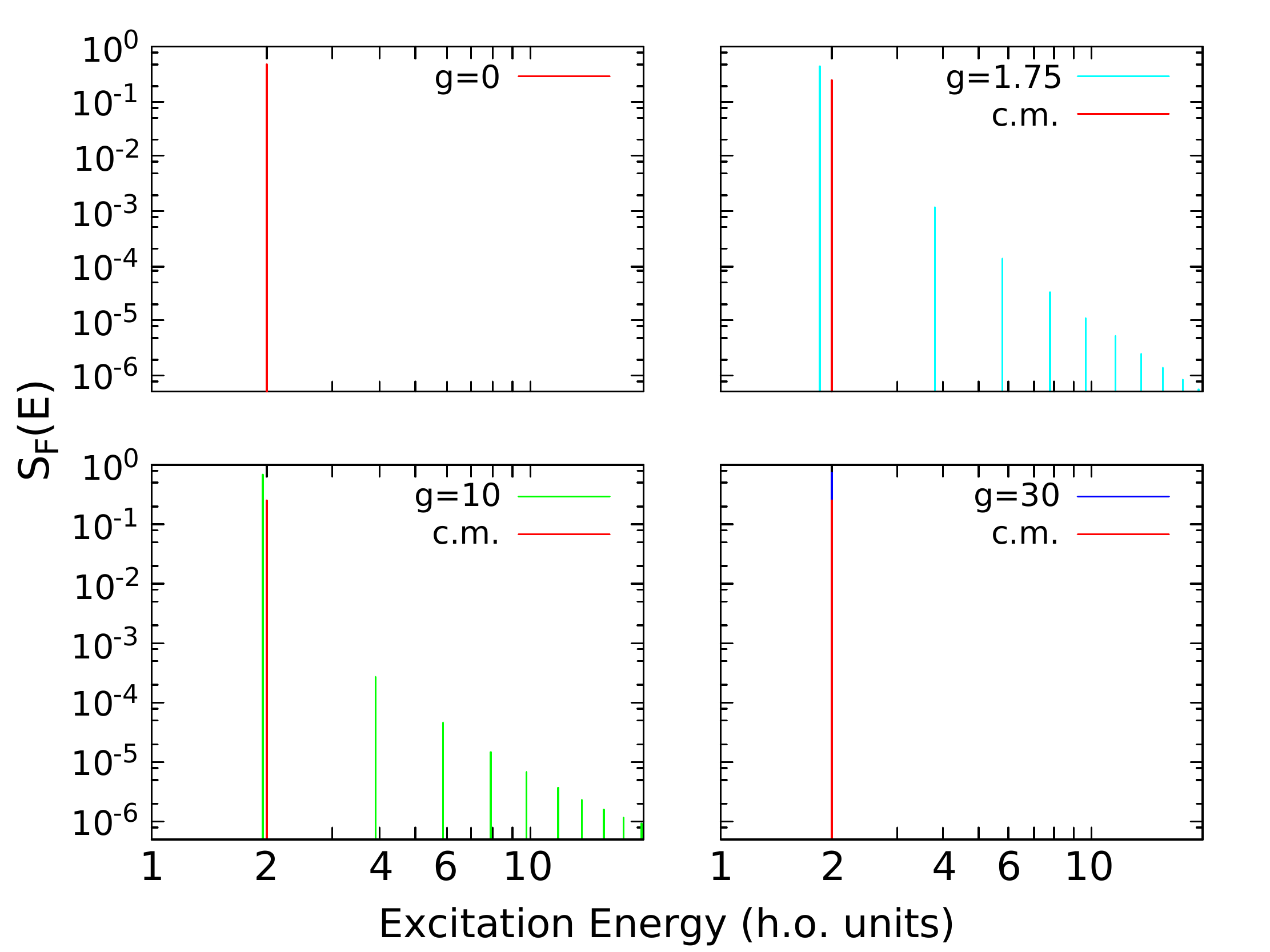}
\caption{Dynamic structure function computed for a system composed by 
2 bosons, with a delta contact potential. The value is calculated for 
$g=0,~1.75,~10,~30$ and with a number of modes of $M=100$.}
\label{fig:fig1}
\end{figure}

The dynamic structure function encodes the response of our system 
to an external perturbation. In this section we will consider the 
dynamic structure function of a monopolar excitation, also known as the 
breathing mode.
For a system with an arbitrary number of particles $N$, the breathing mode can 
be excited by the one body operator $\hat{F}=\sum_{i=1}^{N}x_i^2$. 
Its associated dynamic structure function reads,
\begin{equation}
S_F(E)=\frac{1}{N}\sum_{q}^D{|\langle q|\hat{F}|0\rangle|^2 \ \delta[E-(E_q-E_0)]},
\label{Sf}
\end{equation}
in which $D$ is the dimension of the truncated space of the diagonalization and 
where $|q\rangle$ runs over the excited states, $H |q\rangle = E_q |q\rangle$, 
and  $|0\rangle$ is  the many-body ground state. $S_F(E)$ can be explicitly 
computed with the eigenenergies and eigenfunctions of our system. In 
Fig.~\ref{fig:fig1} we depict the strengths of the dynamic structure function 
computed with $N=2$ for four different values of $g$.
Notice that they contain the factor $1/N$ appearing in the definition of 
$S_F(E)$. For $g=0$ and $g\to\infty$ we can compute $S_F(E)$ analytically. 
Notice that the excitation operator $\hat{F}$ can be split in two pieces 
(see appendix~\ref{appa}), one corresponding to the c.m. and the other to the relative motion.
In the case of N=2 it reads: $x_1^2+x_2^2= 2 X^2 + x_r^2/2$. In an experiment, 
separating the contribution from the c.m. from that of the breathing mode seems 
involved. The response of the system will thus easily mix both contributions.

For $g=0$ the two contributions excite two different states, both at 
the same excitation energy $E_{ex}=2$ ($E_{ex}=E_q-E_0$), being $1/4$ the strength 
of the c.m. excitation and $1/4$ the one of the relative motion.

\begin{figure}[t]
\centering
\includegraphics[width=1\columnwidth, angle=-0]{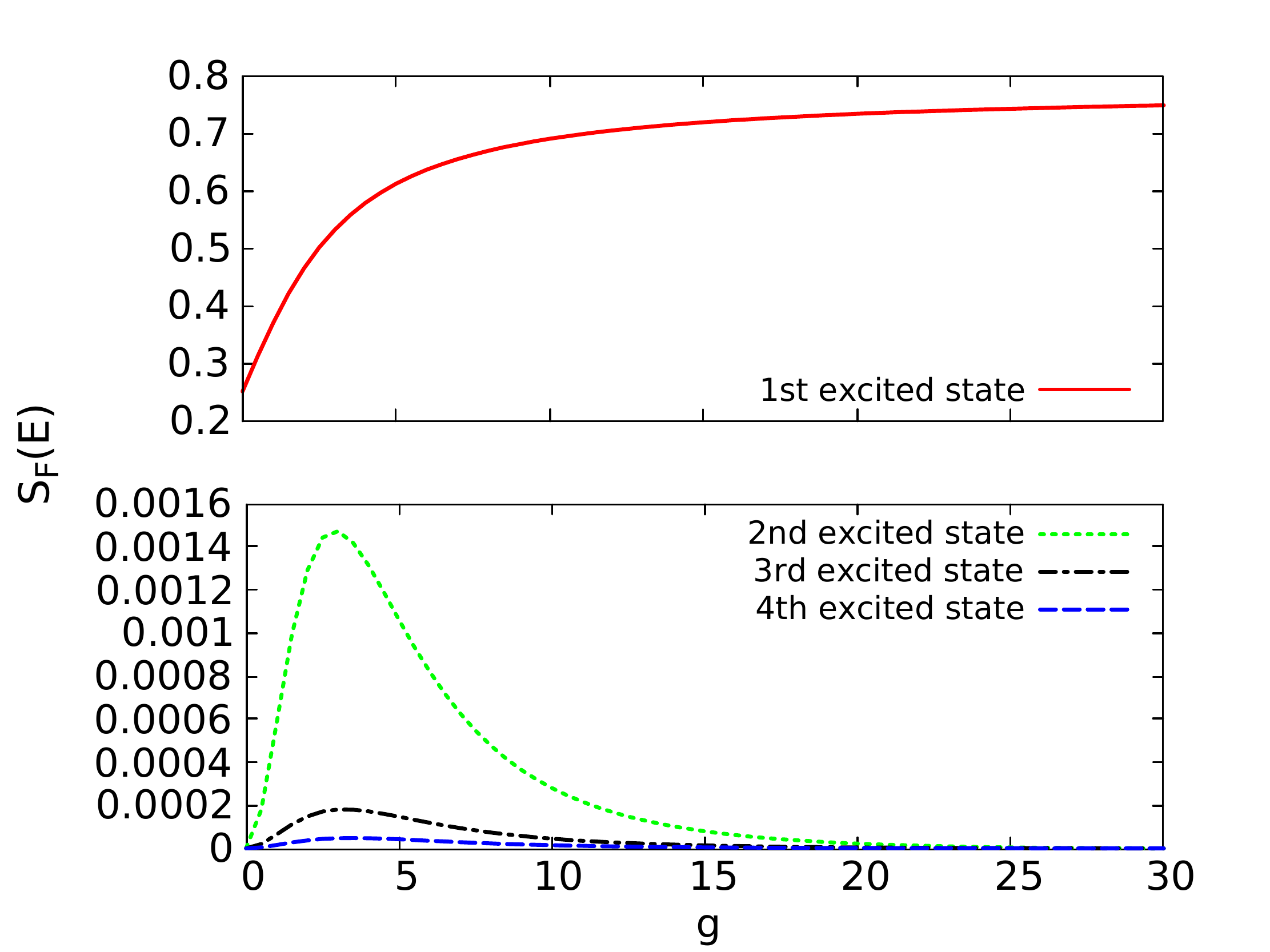}
\caption{Strength of the breathing mode (upper panel) and the next three higher 
relative states excited by the operator $\hat F$ (lower panel) as a function of 
the interaction strength $g$. They contain the factor $1/N$.} 
\label{fig:strength2}
\end{figure}

For nonzero values of $g$ the DSF is distributed over several excited states.
As explained in the appendix A, the c.m. excitation generates only a peak at 
$E_{ex}=2$ with strength $1/(2N)=1/4$, independently of the value of $g$.
The strength of the breathing mode grows monotonously from $1/4$
to $3/4$ as $g$ increases from 0 to infinity. This implies that even though both 
excitations have exactly the same energy it should be more likely to excite 
the breathing mode in the TG limit than in the BEC one.
On the other hand, its excitation energy decreases for increasing values of $g$, up to $g\lesssim 2$.
Then, in the $g\to\infty$ limit, it reaches again the excitation energy with value $E_{ex}=2$.
This reentrant behaviour of the monopolar excitation is reported in Ref.~\cite{breathing}.
The strength of the other peaks are always significantly smaller than those of the c.m. and the breathing mode.
In Fig.~\ref{fig:strength2} we report  the strength of the breathing mode as a function of $g$ and the strength of the next three excited states of the relative motion.
The strength of the breathing mode turns out to be an increasing function of $g$. This behaviour will be relevant to understand the evolution of the energy weighted sum rules of $S_F(E)$.
In all the other cases, the strength is zero for $g=0$ and increases with $g$ up to a maximum, located around $g\sim 2$, to decrease again to zero as $g\to\infty$.
The maximum decreases when the excitation energy increases. Apparently, this dependence is very much 
correlated to the dependence of the interaction energy with $g$ (Fig.~\ref{fig:fign}). 

\begin{figure}[t]
\centering
\includegraphics[width=\columnwidth]{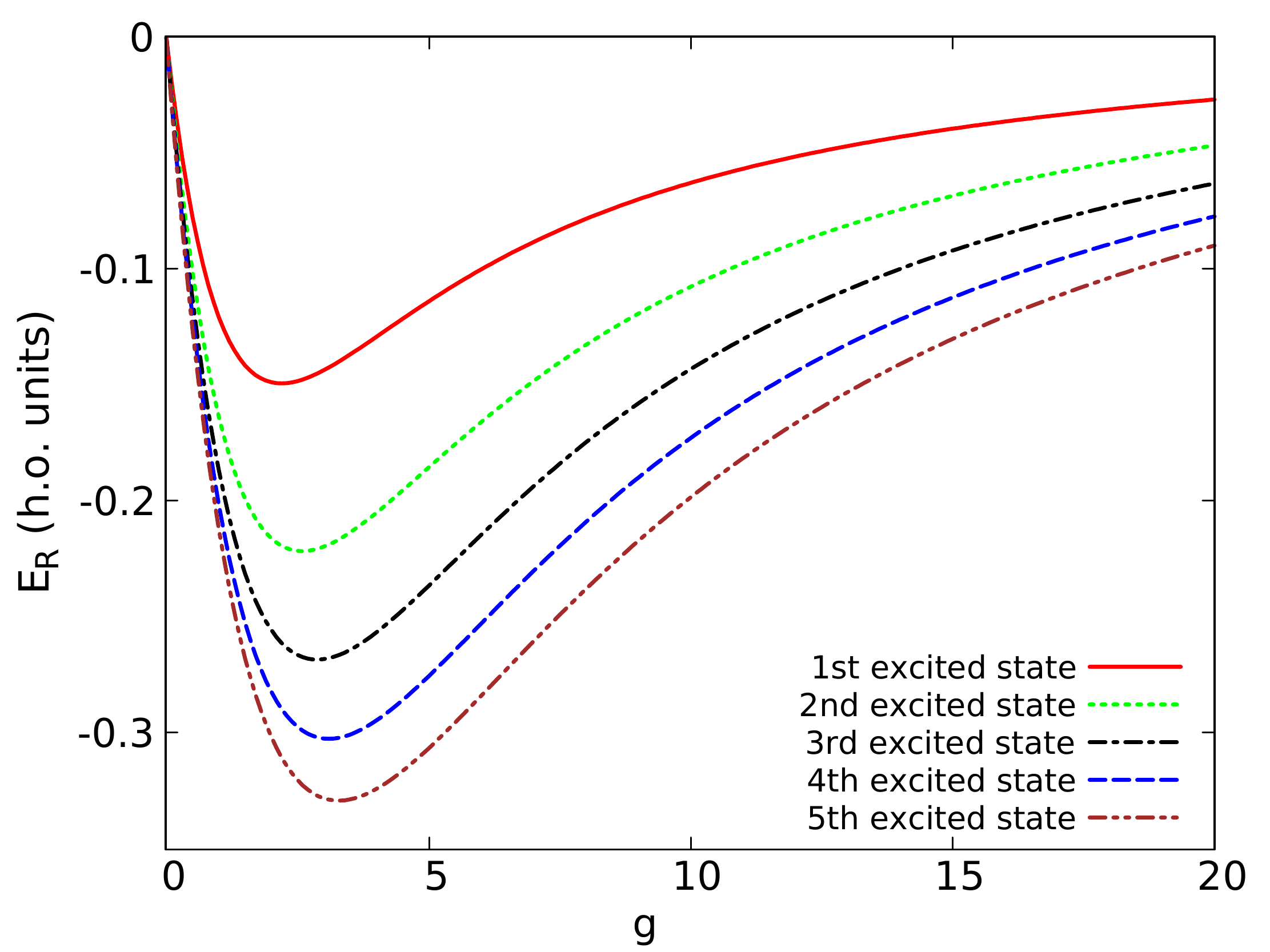}
\caption{Reentrance energy, $E_R=E_{ex}(g)-E_{ex}(g=0)$ as a function 
of the interaction strength $g$ for several excited states of $H_r$.}  
\label{fig:reen}
\end{figure}

The reentrance behavior affects not only the ground state but also all excited states of the relative motion.
As shown in Fig.~\ref{fig:reen} for several excited states, the reentrance 
energy, defined as $E_R=E_{ex}(g)-E_{ex}(g=0)$, is zero by definition at $g=0$.
For increasing values of $g$ it goes through a minimum to increase again 
asymptotically to zero when $g\rightarrow\infty$. This reentrance behaviour 
is more pronounced for the higher excited states. 
The maximum reentrance slightly shifts to higher values of $g$ for the higher 
excited states. Again, there seems to exist a  correlation between the 
interaction energy and the reentrance energy as a function of $g$.
 
Computing the structure function explicitly becomes difficult for more 
than 2 particles. In particular, a naive second quantization scheme using 
the one particle modes as single particle states runs into difficulties 
as it mixes the c.m. with the relative motion in a non-trivial 
way~\cite{schmelcher}. A way to circumvent this problem would be to 
consider the excitation operator, $\hat{\tilde{F}}= \sum_i (x_i -X)^2$, as 
in~\cite{breathing}.

\section{Sum rules}
\label{sumrules}

An alternative method to characterize the dynamic structure function, 
$S_F(E)$, is by calculating its momenta $M_n$,
\begin{eqnarray}
M_n&=&\int_0^{\infty}E^n \ S_F(E)\ dE \nonumber\\
& =& 
\frac{1}{N}\sum_q^D(E_q-E_0)^n  \ |\langle q|\hat{F}|0\rangle|^2\,.
\label{Mn}
\end{eqnarray}
As it is well known, there are some theorems that allow one to 
compute several of the $M_n$ without any explicit knowledge of all 
the eigenvalues and eigenvectors of the many-body system~\cite{joan}. 
This is done through the so-called sum rules. These sum rules will be 
computed by suitable operations on the ground state of the system. 

Also note that the momenta $n$ and $n+2$ provide estimates for the 
monopolar excitation energy which is how Monte-Carlo calculations 
proceed~\cite{breathing}. Assuming the structure function  concentrated 
around a single excitation energy, $S_F\simeq S_m \delta(E-E_m)$, then 
we have $E_m \simeq \sqrt{M_n/M_{n-2}}$ for all $n$. The simplest one would be 
$E_m\simeq \sqrt{M_1/M_{-1}}$.
The sum rules below are derived in full generality, independent of the number of particles of the many body system and, thus, they are particularly useful combined with Monte-Carlo techniques which 
are able to describe the ground state. However, all the calculations reported 
in the paper are for $N=2$ and have been performed taking into account 
the separation of the c.m. and the relative motion. The contributions 
of the c.m. to the considered sum rules are all analytical. 
\begin{figure}[t]
\centering
\includegraphics[width=1\columnwidth, angle=0]{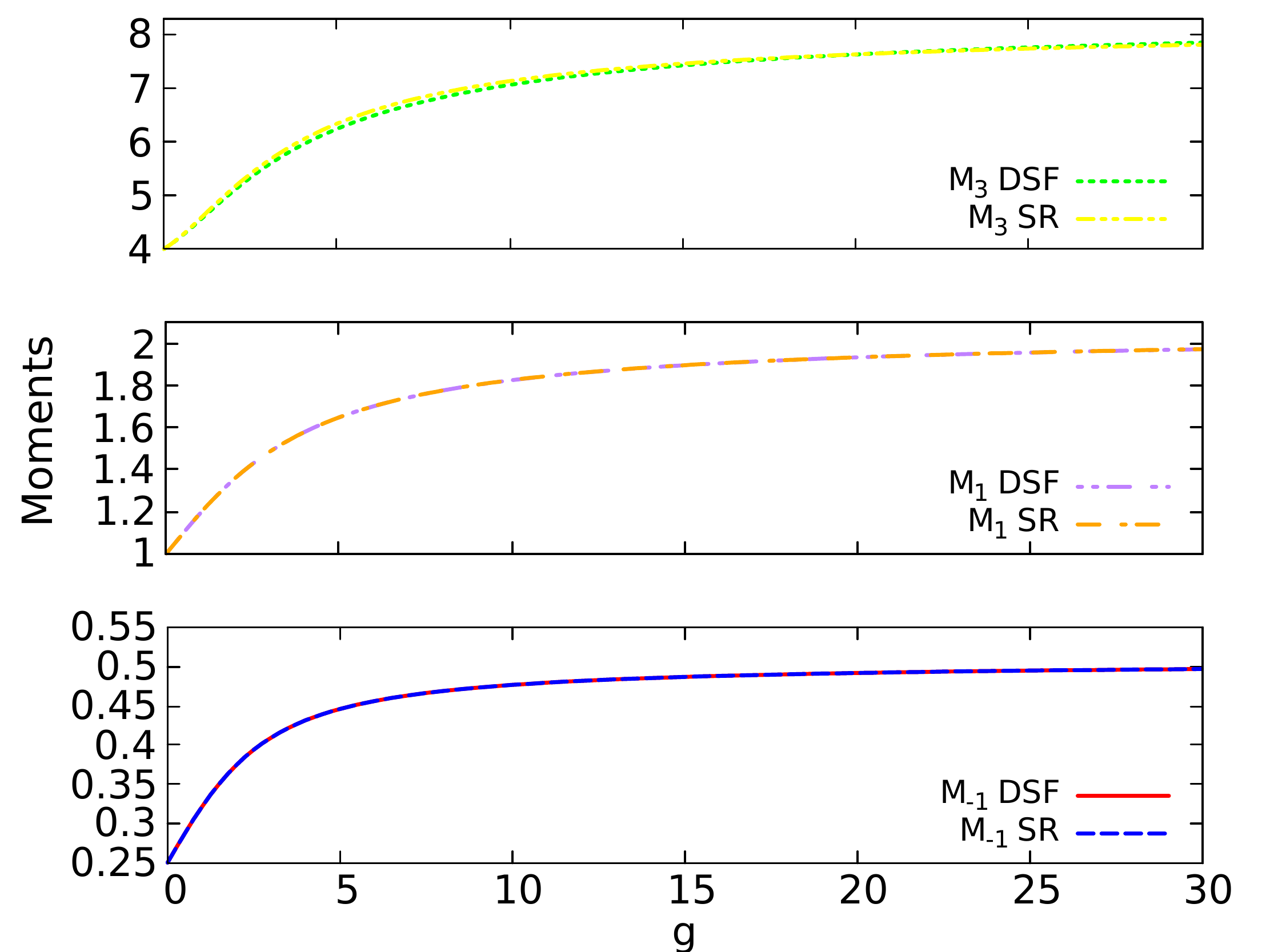}
\caption{$M_{-1}$, $M_1$ and $M_3$ computed directly from $S_F(E)$ (DSF) and 
the sum rules (SR). When using $S_F(E)$ we 
have truncated the spectra by taking the first 10 energy levels.}
\label{fig:fig2}
\end{figure}

\subsection{$M_{-1}$ sum rule}

To obtain the  $M_{-1}$ sum rule, we build a new Hamiltonian 
$\tilde{H}=\hat{H}+\lambda\hat{F}$ and consider the new term as a 
perturbation modulated by the  parameter $\lambda$. The expansion 
up to second order of the ground state energy $E_0(\lambda)$, 
eigenenergy of $\tilde{H}$, reads,
\begin{equation}
E_0(\lambda)=E_0+
\lambda\langle 0|\hat{F}|0\rangle+\lambda^2
\sum_q\frac{|\langle q|\hat{F}|0\rangle|^2}{E_0-E_q}\,,
\label{E_0(lambda)}
\end{equation}
where $|q\rangle$ and $E_q$ are eigenstates and eigenvalues 
of the unperturbed Hamiltonian, respectively. 
Therefore, $M_{-1}$ can be written  as, 
\begin{equation}
M_{-1}=-\frac{1}{2}\frac{1}{N}
\left.\frac{\partial^2E_0(\lambda)}{\partial\lambda^2}\right| _{\lambda=0}\,.
\label{M_-1 E0}
\end{equation}
Using also perturbation theory, we have an alternative  expression  for $M_{-1}$,
\begin{equation}
M_{-1}=
-\frac{1}{2}\frac{1}{N}\left.\frac{\partial}{\partial\lambda}\langle 
\tilde{0}|\hat{F}|\tilde{0}\rangle\right| _{\lambda=0}\,,
\label{M_-1 F}
\end{equation}
where $|\tilde{0}\rangle$ is the ground state of the perturbed Hamiltonian 
$\tilde{H}$. For the present monopole operator,
\begin{equation}
\frac{1}{N}\langle \tilde{0}|\hat{F}|\tilde{0}\rangle=
\frac {1}{N} \langle \tilde{0}|\sum_i x_i ^2|\tilde{0}\rangle
= \int_{-\infty}^{\infty}x^2\rho_\lambda(x)dx \,, 
\end{equation}
with $\int_{-\infty}^{\infty}\rho_\lambda(x)dx=1$.

\subsection{$M_1$ sum rule}

The $M_1$ sum rule  can be written as, 
\begin{eqnarray}
M_1&=&\frac{1}{2}\frac{1}{N}
\langle0|[\hat{F}^\dagger,[\hat{F},\hat{H}]]|0\rangle 
= 2 \frac {1}{N} \langle 0|\sum_i x_i^2|0\rangle \nonumber \\
&=& 2 \int_{-\infty}^{\infty}x^2\rho(x)dx = \frac {4}{N} \ V_{\rm h.o.}\,,
\label{M_1}
\end{eqnarray}
with $\int_{-\infty}^{\infty}\rho(x)dx=1$ and $V_{\rm h.o.}$ is the total harmonic 
potential energy. 

In Ref.~\cite{breathing} the monopolar excitation energy is estimated by 
means of $E\simeq \sqrt{M_1/M_{-1}}$ with $M_{-1}$ obtained from Eq.~(\ref{M_-1 F}). 

\subsection{$M_3$ sum rule}

Assuming that $\hat{F}$ is hermitian, $M_3$ can be calculated as the following 
expectation value:
\begin{equation}
M_3 = \frac {1}{2N} 
\langle 0 | 
\left [ [\hat{H},\hat{F}], \left [\hat{H}, [\hat{H},\hat{F}] \right ] \right ] 
| 0 \rangle \, .
\end{equation}
Alternatively, it can also be calculated as 
\begin{equation}
M_3 = \left.\frac {1}{2N} \frac {\partial ^2 E_{\eta}}{\partial \eta^2}  \right|_{\eta =0} \, ,
\end{equation}
where
\begin{equation}
E_{\eta} = \langle \phi_{\eta}| \hat{H} |\phi_{\eta} \rangle \;{\rm with}\;
| \phi_{\eta} \rangle = e^{\eta [\hat{H},\hat{F}]} | 0 \rangle\,.
\end{equation}
By means of the  virial theorem derived from, 
\begin{eqnarray}
E_{\lambda} &=& \langle \Psi_{\lambda} |\hat{H} | \Psi_{\lambda}\rangle\\ 
&=&
\lambda^2 ~E_{\rm kin}^{\lambda=1}
-\frac{1}{\lambda^2} V_{\rm h.o.}^{\lambda=1} + \lambda~ V_{\rm int}^{\lambda=1} \nonumber \, ,
\end{eqnarray}
 with 
\begin{equation} 
\Psi_{\lambda}(x_1,x_2,\dots,x_N) \equiv \lambda^{N/2} \Psi(\lambda x_1,\lambda x_2,\dots,\lambda x_N)\,,
\end{equation}
and noting that 
\begin{equation}
e^{\eta [H,F]} \Psi(x_1,x_2,\dots,x_N) =
\Psi(e^{\eta} x_1,e^{\eta} x_2,\dots ,e^{\eta} x_N) \,,
\end{equation}
$M_3$ can then be computed as,
\beq
M_3 = \frac {4}{N} (E_{\rm kin} + 3 V_{\rm h.o.}) \,.
\eeq

\subsection{Behaviour and asymptotic values of the sum rules}

\begin{figure}[t]
\centering
\includegraphics[width=1\columnwidth, angle=-0]{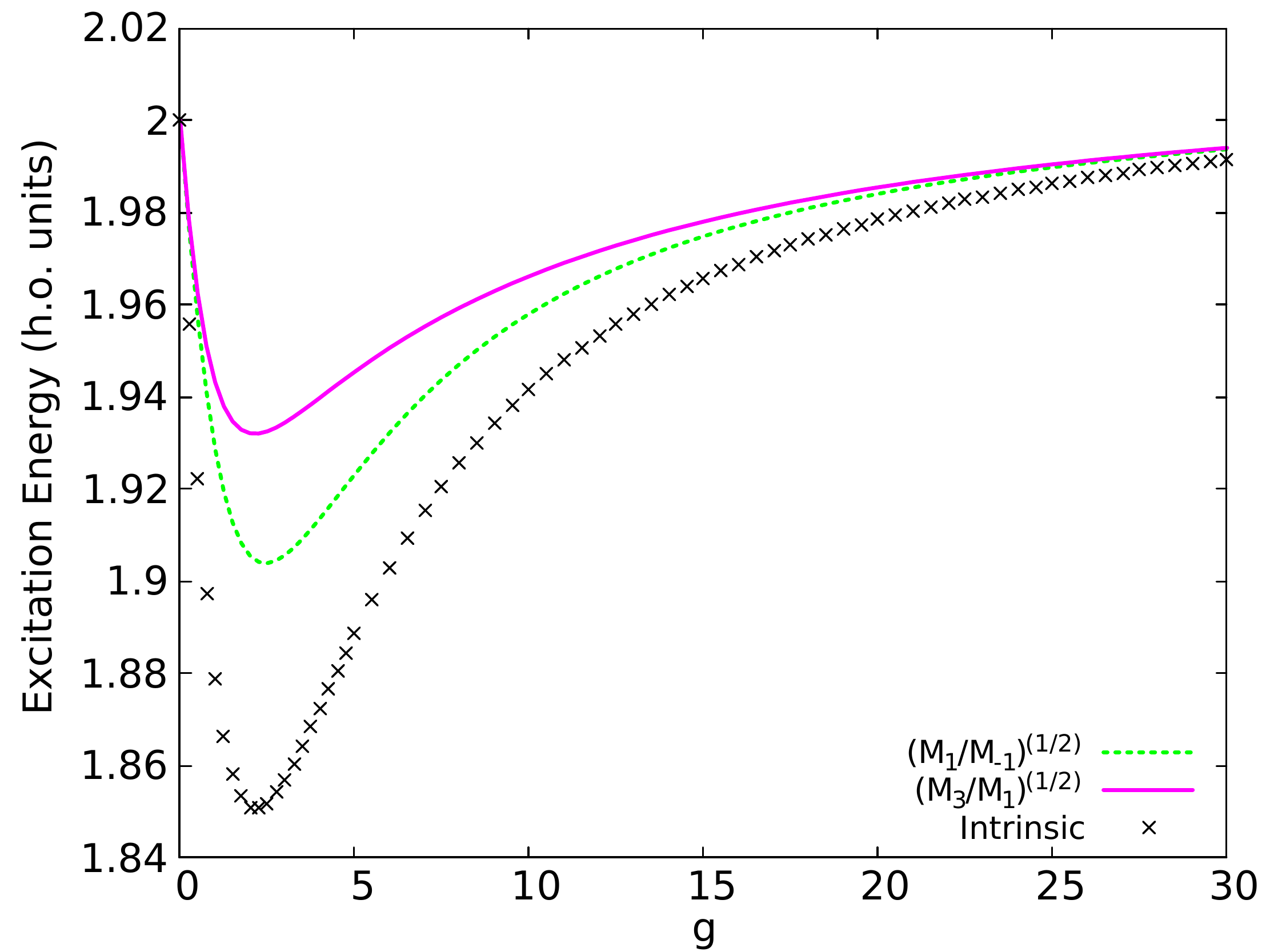}
\caption{Monopolar excitation energies obtained through $M_{-1}$, $M_1$ and $M_3$, 
and first excitation energy from the spectra.} 
\label{fig:fig3}
\end{figure} 

In Fig.~\ref{fig:fig2}  the sum rules $M_3$, $M_1$ and $M_{-1}$ are reported as a 
function of $g$. The sum rules have been calculated by 
explicitly integrating $S_F(E)$ or by using the formulas of the sum rules in terms 
of expectation values of certain operators in the ground state. The agreement between 
both methods is excellent in the full range of $g$ explored. 
All the sum rules are increasing functions of the interaction strength and approach an 
asymptotic value for $g \rightarrow \infty$. The increasing character of the sum rules 
with $g$ is mainly due to the increment of the strength of the breathing mode with 
$g$.
As expected, the convergence to the asymptotic values is faster for $M_{-1}$ and 
slower for $M_3$.
However, all cases have almost reached convergence already for $g=30$.

As explained in the appendix~\ref{appa}, due to the  operator $x_1^2+x_2^2$ the only excitation of the c.m. lies at an energy $E_{ex}=2$, above the ground state.
The strength of this excitation (which contains the factor $1/N$) is $1/4$ and therefore 
the contribution to the different sum rules of the c.m. excitation is given by 
\begin{equation} 
	M_{-1,c.m.} = \frac {1}{8} ~~~,~~~ M_{1,c.m.} = \frac {1}{2}~~~,~~~ M_{3,c.m.} = 2 \, .
\end{equation}
The contributions to the different sum rules of both the energy and the 
strength of the c.m. excitation are independent of $g$.

The non-interacting and the infinitely interacting limits are particularly amenable 
and allow one to compute the exact limiting values of the $M_1$ and $M_3$ sum rules. 
The two sum rules can be cast as a function of the average values of the kinetic 
and harmonic oscillator energies. The $M_1$ and $M_3$ sum rules for $N$ particles read,
\beqa
M_1 &=& \frac {4}{N} V_{\rm h.o.} \nonumber \\
M_3 &=& \frac {4}{N} (E_{\rm kin} + 3 V_{\rm h.o.})   
\eeqa
where $V_{\rm h.o.}$ is the expectation value of the total harmonic oscillator 
potential energy in the $N$-particle wave function and $E_{\rm kin}$ is the 
total kinetic energy associated to the ground state.

\begin{figure}[t]
\centering
\includegraphics[width=1\columnwidth, angle=-0]{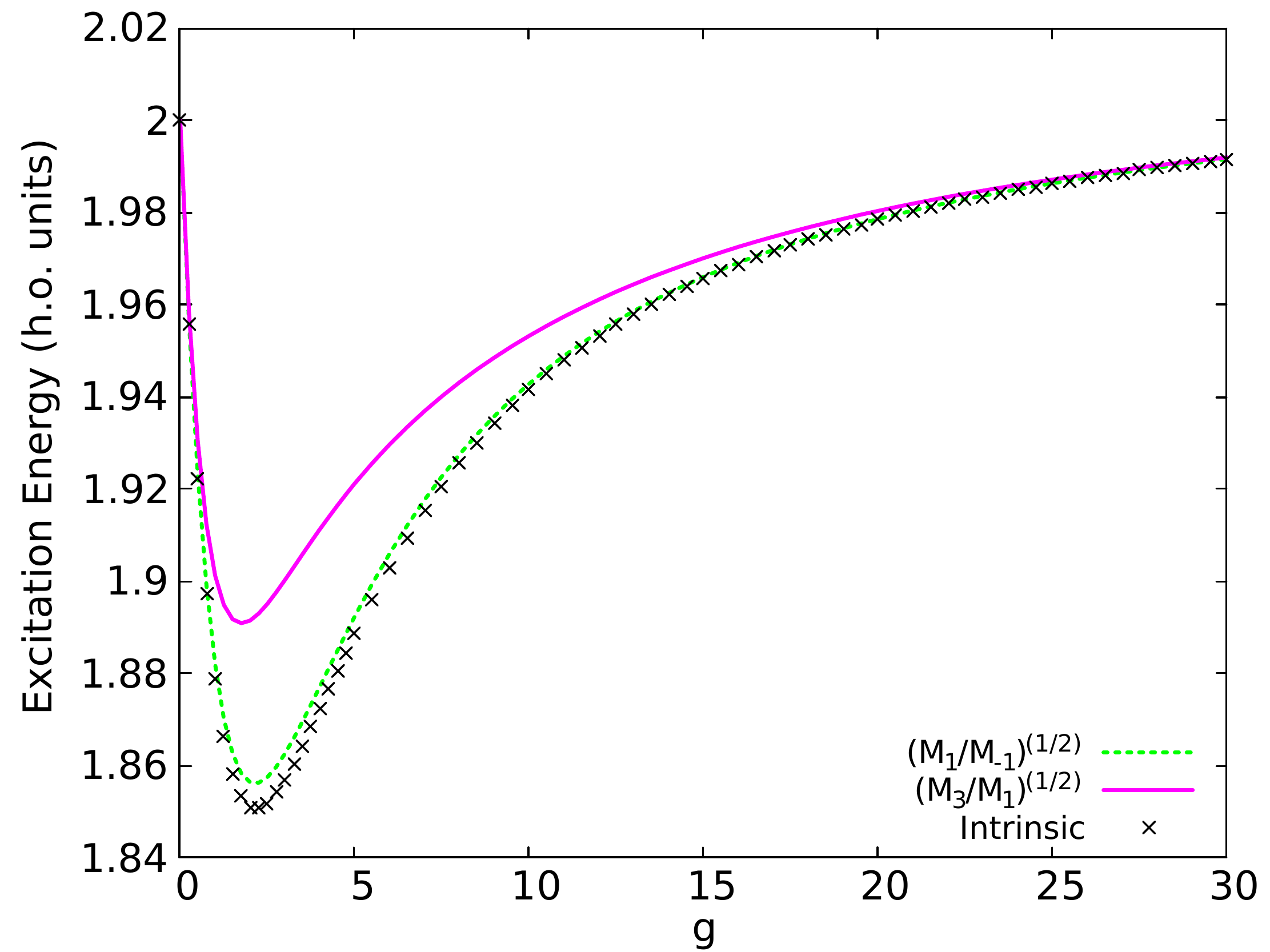}
\caption{Monopolar excitation energies obtained through $M_{-1}$, $M_1$ and $M_3$, 
and the energy of the breathing mode. The contribution of the c.m. 
to the sum rules has been subtracted.} 
\label{fig:fig4}
\end{figure} 
For $g=0$, all particles populate the h.o. single-particle ground state.
Therefore, the total, kinetic and harmonic potential energies are 
$E = N /2 $, $E_{\rm kin} =  N /4$, and $V_{\rm h.o.} = N/4$, respectively.
Hence, for $N=2$ we have, $M_3 = 4$ and $M_1= 1$ and  the estimate of the monopolar excitation 
energy characterizing the dynamic structure function $S_F(E)$ is
\begin{equation}
E_{\rm ex}= {\sqrt { \frac {M_3}{M_1} }} = 2 \,.
\end{equation}

To evaluate the sum rules in the limit $g \rightarrow \infty$ we rely on 
the Bose-Fermi mapping. In this case, the ground state is the absolute value 
of the Slater determinant built as if the particles were fermions. 
In the case of $N$ particles they occupy the first $N$ lowest 
single-particle energy levels of the harmonic oscillator potential so that
\begin{equation}
E =  \left(\frac {1}{2} + \frac {3}{2} + \dots
+ \frac {2 N -1}{2}\right)   
= \frac {1}{2} N^2 
\end{equation}
while $E_{\rm kin} = N^2/4$, and $V_{\rm h.o.} = N^2/4$.

Consequently,
\begin{equation}
M_1 = \frac {4}{N}  V_{\rm h.o.} = N  \,,\; M_3  
= \frac {4}{N} (E_{\rm kin} + 3 V_{\rm h.o.})  = 4 N  
\end{equation}
which for $N=2$ results in $M_1= 2$ and $M_3 = 8$.
then, the monopolar excitation energy calculated with $M_3$ and $M_1$ is, in this case,
\begin{equation}
E_{\rm ex}= {\sqrt { \frac {M_3}{M_1} }}= {\sqrt {\frac { 4 N}{N}}}= 2 \, ,
\end{equation}
which is the same as in the non-interacting case. 

Recalling to Sec.\ref{dynstrufunc}, the intrinsic motion energy and strength 
have a dependence on $g$, contrary to those of the c.m. excitation. In particular, 
the strength is an increasing function of $g$ that goes from 1/4 at $g=0$ to 
3/4 at $g\to\infty$. This increment of the strength of the intrinsic peak 
associated to the monopole vibration is the main reason for the increasing behaviour of 
the different sum rules as a function of $g$. For all values of $g$, the 
peaks associated to the c.m.  and to the breathing mode exhaust more than 
$99 \%$ of the sum rules and the other peaks have a marginal influence. 

In 1D ultracold gases with atom-atom contact interaction is sensible to neglect 
the interaction energy in front of the kinetic energy when the interaction 
strength is increased, unlike in a Thomas-Fermi limit where the kinetic energy 
is neglected in front of the interaction energy. Therefore, it is also illustrative to compare the $g\to \infty$ value with the Thomas-Fermi 
one, where the kinetic energy is neglected and the estimate of the excitation 
energy from the sum rules is $E_{\rm ex}=\sqrt{3}$. Thus, as emphasized in~\cite{breathing}, 
the reentrant behavior is a clear consequence of the fermionization taking place 
as $g\to\infty$. In $1D$ gases with contact interaction, as the interaction 
is increased the approximation which is sensible is to neglect the interaction 
energy in front of the kinetic one, unlike in the Thomas-Fermi limit where the 
kinetic energy is neglected in front of the interaction. 

As the important peaks (c.m. and breathing mode ) of $S_F(E)$ are rather close 
to each other one expects the estimation of the energy through the sum rules 
to be rather accurate. In Fig.~\ref{fig:fig3} we compare the monopolar excitation 
energies estimated by means of the sum rules described above with the values of 
the excitation energy of the breathing mode, $E_{0,2}-E_{0,0}$.  Notice that in all 
cases, the estimate $\sqrt {M_3/M_1}$ is larger than $\sqrt {M_1/M_{-1}}$ for any 
value of $g$ and both are larger or equal than the minimum excitation energy 
defined by $E_{0,2} - E_{0,0}$. First, we note that the two quantities  obtained 
from $\sqrt{M_1/M_{-1}}$ and $\sqrt{M_3/M_1}$ do not produce the same estimate 
for the intrinsic monopolar excitation energy. This reflects the fact that the 
structure function contains two relevant excited states  not located at the same 
energy, one from the c.m. and the second from the breathing mode. The 
difference between the two estimates provided by the sum rules increases with 
$g$ and is larger for the value of $g$ that maximizes the interaction energy. 
For this value of $g$ the difference in energy between the c.m. excitation and 
the energy of the breathing mode is also maximal. Then, when $g$ increases further 
and the breathing mode gets closer again to the c.m. excitation, the estimates 
get closer and both reach asymptotically the $E_{ex} =2$ when $g \rightarrow \infty $. 
Subtracting the contribution of the c.m. to the sum rules, which is equivalent to 
study only the the intrinsic excitations (see Fig. ~\ref{fig:fig4}), there is only 
one dominant peak: the breathing mode, and the estimations from the sum rules 
get much closer to each other. The differences in this case should be assigned to 
the higher excitations beyond the breathing mode.

\section{Summary and Conclusions} 
\label{conclusions}

We have considered a system of two bosons trapped in a 1D harmonic oscillator 
potential. The interaction between the two bosons is assumed to be well represented 
by a contact term. Under these assumptions we have computed the dynamic structure 
function for a monopolar excitation as a function of the interaction strength. Our 
method consists in diagonalizing the Hamiltonian of the relative motion in a 
large enough truncated basis. In this way we are able to obtain stable results 
for both the positions and strengths of the peaks of the dynamic structure function. 
We have compared the direct calculation of the lower momenta of the 
dynamic structure function with computations performed by means of sum rules. 
The agreement between both methods is almost perfect for $M_{-1}$ and $M_1$, and 
is within $2\%$ check this percentage for $M_3$. We have computed two 
estimates of the monopolar excitation energy obtained from $\sqrt{M_{1}/M_{-1}}$ and 
$\sqrt{M_{3}/M_{1}}$. Differences between both estimates reflect the fact 
that a minimum of two peaks (c.m. and breathing mode) are always contributing to 
the structure function for any value of $g$. When the c.m. contribution is subtracted 
from the sum rules, the prediction of the excitation energy of the intrinsic 
breathing mode improves substantially. Finally, as shown in the appendix, 
the contribution of the c.m. to the sum rules decreases with the number of particles. 
Therefore, it is expected that the sum rule approach to estimate the monopole 
excitation will work better for a larger number of particles. 

\begin{acknowledgments}
The authors acknowledge useful discussions with Joan Martorell and Miguel 
Angel Garc\'ia March. We acknowledge  financial support from the Spanish 
Ministerio de Economia y Competitividad Grant No FIS2017-87534-P, 
and from Generalitat de Catalunya Grant No. 2014SGR401.  
\end{acknowledgments}

\clearpage

\appendix

\section{Center-of-mass for N-particles and its monopolar excitation}
\label{appa}

An interesting feature of this  two-body problem is the relation between the 
many-body correlations and the monopolar excitation in the system. In any 
experimental implementation it will be difficult, if not impossible, 
to separate the center-of-mass (c.m.) contribution from the internal one, which is 
the one carrying information about the correlations. Notably, for an arbitrary 
number of particles, the c.m. contribution can be evaluated analytically. 

Lets start from the following identity, 
\begin{equation}
\sum_{i=1}^n x_i^2 = N X^2 + \frac {1}{N} \sum_{i<j} (x_i- x_j)^2 \, .
\end{equation}
and consider an $N$-particle Hamiltonian with interactions depending on 
the interparticle distance, confined in a harmonic oscillator potential,
\begin{equation}
H= - \sum_{i=1}^N \frac {1}{2} \frac {d^2}{dx^2} 
+ \frac {1}{2} \sum_{i=1}^N x_i^2 + \sum_{i<j} v(x_i-x_j)  \,.
\end{equation}
Which can be decomposed in two pieces,
\begin{equation}
H = H_{\rm c.m.}+ H_{r}
\end{equation}
where $H_{\rm c.m.}$:
\begin{equation}
H_{\rm c.m.} = - \frac {1}{2 N} \frac {d^2 }{dX^2}  + \frac {N}{2} X^2
\end{equation}
describes the c.m. motion and $H_{r}$ affects only 
the relative coordinates and is translationally invariant. 
Notice that for $N=2$, we recover the c.m. Hamiltonian of the 
two particle case, Eq.~(\ref{eq:3}). Therefore the many-body wave functions of the system can be 
factorized,
\begin{equation}
\Psi_j (x_1,x_2, ...,x_n) = \varphi_{j_{\rm c.m.}} (X) \phi_{j_{r}} (\{ x_i^{r} \} ) \, .
\end{equation}

Let's now go back to the definition of the dynamic structure function: 
\begin{equation}
S_F(E) = \frac {1}{N} \sum_j \bigg| \langle j | 
\sum_{i=1}^N x_i^2 | 0 \rangle \bigg| ^2 \delta (E-(E_j-E_0)) \, ,
\end{equation}
which can also be separated in a piece corresponding to the 
center-of-mass an another piece affecting only the relative coordinates:
\beqa
&&\langle j | \sum_i x_i^2 | 0 \rangle = \\
&&\langle j_{\rm c.m.}, j_{r} | 
N X^2 + \frac {1}{N} \sum_{i<j} (x_i-x_j)^2 | 0_{\rm c.m.},0_{r} \rangle  \nonumber
\eeqa
where  $| 0,0\rangle = \varphi_0^{\rm c.m.}(X) \phi_0^{r}(\left[x_i^{r}\right ]) $, 
\beqa
\langle j | \sum x_i^2 | 0 \rangle  &=& 
\langle j_{\rm c.m.}| N X^2 | 0_{\rm c.m.} \rangle \langle j_{r} | 0_{r}\rangle   \\
 &+& \langle j_{\rm c.m.} | 0_{\rm c.m.} \rangle  \langle j_{r} | \frac {1}{N} \sum_{i<j} (x_i-x_j)^2 
| 0_{r} \rangle \, . \nonumber 
\eeqa

Taking into account the orthogonality of the states, 
\beqa
\langle j | \sum x_i^2 | 0 \rangle &=& \langle j_{\rm c.m.}| N X^2 | 0_{\rm c.m.} \rangle
\delta _{j_{r}, 0_{r}}   \\
&+& \delta_{j_{\rm c.m.}, 0_{\rm c.m.}} \langle j_{r} | \frac {1}{N} \sum_{i<j} (x_i-x_j)^2 | 0_{r} \rangle \, .
\nonumber
\eeqa
Therefore, we  have separated the excitations of the $\rm c.m.$ from the intrinsic ones. 
Now, we should take into account that the spectrum of the $\rm c.m.$ is indentical to the 
harmonic oscillator: $1/2, 3/2, 5/2 ,....$ and that the ground state will correspond always 
to the $j_{\rm c.m.}=0$ state with energy $1/2$. With this excitation operator, the only 
excitation of the $\rm c.m.$ will correspond always to excite the state $j_{\rm c.m.}= 2$ with 
energy $E^{\rm c.m.}= 5/2$,  with an excitation energy $E_{exc}^{\rm c.m.}= 2$. The matrix element, 
for $j_{\rm c.m.} \ne 0$  fulfils that: 
\begin{equation}
\langle j_{\rm c.m.} | N X^2 | 0_{\rm c.m.} \rangle = \delta_{j_{\rm c.m.}, 2} N \langle 2_{\rm c.m.} | X^2 | 0_{\rm c.m.} \rangle \, .
\end{equation}
Actually, one can evaluate this  matrix element:
\begin{equation}
\langle 2_{\rm c.m.} | X^2 | 0_{\rm c.m.} \rangle = \frac {1}{{\sqrt{2}} N}
\end{equation}
and 
\begin{equation}
\big| \langle 2_{\rm c.m.} | N X^2 | 0_{\rm c.m.} \rangle \big|^2 = \bigg| N \frac {1}{{\sqrt{2}} N} \bigg|^2 = \frac {1}{2} \, .
\end{equation}
Therefore, the c.m. excitation appears always at an excitation energy $E_{ex}=2$ with 
strength $1/(2N)$, independently of the  interactions between the particles. The factor 
$1/N$ is due to our definition of $S_F(E)$.

\subsection{ Contribution of the {\rm {c.m.}} excitations to the different sum rules}

We have shown that the only excitation of the $\rm c.m.$ due to $F= \sum x_i^2$ 
lies at an energy $E_{ex} = 2$, above the ground state, independently of the 
interaction between the particles. The strength of this excitation is given by $1/2N$.  
Contributions of this $\rm c.m.$ excitation to the energy weighted sum rules: 
\begin{eqnarray}
M_{-1,\rm c.m.}& =& \frac {1}{E_{ex}} \frac {1}{N} \frac {1}{2} = \frac {1}{4} \frac {1}{N} \nonumber \\
M_{1,\rm c.m.} &=& E_{ex} \frac {1}{N} \frac {1}{2} =\frac {1}{N}  \nonumber \\
M_{3,\rm c.m.} &=& E_{ex}^3 \frac {1}{N} \frac {1}{2} = \frac {4}{N}
\end{eqnarray}
that for $N=2$ become: 
\begin{equation}
 M_{-1,\rm c.m.}= \frac {1}{8} ~~~~,~~~
M_{1,\rm c.m.} =\frac {1}{2} ~~~~,~~~ M_{3,\rm c.m.}= 2.
\end{equation}

These results  can be always used as a  test for the calculations. 

\end{document}